\documentclass[english,showpacs,preprint,superscriptaddress]{revtex4-1}

 \UseRawInputEncoding  
 
\usepackage{babel}
\usepackage[T1]{fontenc}      
\usepackage{mathrsfs}         
\usepackage{bm}
\usepackage{color, xcolor}   
 
\usepackage{lmodern}     

\usepackage{amsmath,amsfonts,amssymb}
\usepackage{esint}     
\usepackage{extarrows}

\newtheorem{thm}{Theorem A.}

\usepackage[ruled,linesnumbered]{algorithm2e}
\usepackage{algpseudocode}

\usepackage{graphicx}   
\usepackage{float}
\usepackage{tikz}       
\usepackage{palatino}
\usetikzlibrary{arrows,shapes,chains}

\usepackage{subfig}     
\usepackage{booktabs}  
\graphicspath{{papers/All/figure/}}

\usepackage[justification=centering]{caption}
\usepackage{siunitx}

\usepackage{indentfirst}
\usepackage{setspace}

\makeatletter

\usepackage[T1]{fontenc}      

\usepackage[title]{appendix}

\usepackage{hyperref}  
\hypersetup{
    breaklinks = true,
    colorlinks = true,
    citecolor = {blue},
    urlcolor = {blue},
    linkcolor = {blue}
} 

\usepackage{cleveref}

 \global\long\def\FIG#1{~\ref{#1}}
 \global\long\def\EQ#1{~(\ref{#1})}
 \global\long\def\EQo#1{(\ref{#1})}
 
 \global\long\def\EQgive#1{~(give in Eq.~(\ref{#1}))}
 \global\long\def\SEC#1{~\ref{#1}}
 \global\long\def\APP#1{~\ref{#1}}

\makeatother

\begin{document}
 
\title{Relaxation model for a homogeneous plasmas with spherically symmetric velocity space}

\author{Yanpeng Wang}
\affiliation{School of Nuclear Sciences and Technologys, University of Science and Technology of China, Hefei, 230026, China}
\author{Jianyuan Xiao}
\email{Corresponding author. E-mail: xiaojy@ustc.edu.cn}
\affiliation{School of Nuclear Sciences and Technologys, University of Science and Technology of China, Hefei, 230026, China}
\affiliation{Laoshan Laboratory, Qingdao, 266100, China}
\author{Xianhao Rao}
\email{Corresponding author. E-mail: rrxxhh@mail.ustc.edu.cn}
\affiliation{School of Nuclear Sciences and Technologys, University of Science and Technology of China, Hefei, 230026, China}
\author{Pengfei Zhang}
\affiliation{School of Physical Sciences, University of Science and Technology of China, Hefei, 230026, China}
\author{Yolbarsop Adil}
\author{Ge Zhuang}
\affiliation{School of Nuclear Sciences and Technologys, University of Science and Technology of China, Hefei, 230026, China}
  
\begin{abstract}
  We derive the transport equations from the Vlasov-Fokker-Planck equation when the velocity space is spherically symmetric. 
  The Shkarofsky's form of Fokker-Planck-Rosenbluth collision operator is employed in the Vlasov-Fokker-Planck equation. 
  A closed-form relaxation model for homogeneous plasmas could be presented in terms of Gauss hypergeometric2F1 functions. This has been accomplished based on the Maxwellian mixture model.
  Furthermore, we demonstrate that classic models such as two-temperature thermal equilibrium model and thermodynamic equilibrium model are special cases of our relaxation model and the zeroth-order Braginskii heat transfer model can also be derived.
  The present relaxation model is a nonequilibrium model based on the 
  hypothesis that the plasmas system possesses finitely distinguishable independent features, without relying on the conventional near-equilibrium assumption.  
   
   
\end{abstract}

\maketitle



 \UseRawInputEncoding




 
\begin{spacing}{0.35}  

\section{Introduction}
\label{sect:Introduction}

In fusion plasmas, the transport processes between different plasmas species, as well as the wave-particles interactions, are crucial in shaping the evolution of the non-equilibrium and nonlinear plasmas system\cite{Chen1984}. Transport, since it involves collisions, has traditionally been modeled using Vlasov-Fokker-Planck (VFP). However, simulations for these processes usually not only face challenges in conserving mass, momentum, and energy in discrete\cite{Thomas2012}, but also satisfying higher-order moment convergence\cite{wang2024Aconservative,wang2024General} to effectively capture the inherent nonlinearity of plasmas system. 
Additional difficulties arise from the significant differences in thermal velocities due to mass or energy discrepancies\cite{wang2024Aconservative,wang2024General,Taitano2016} in collisions, which typically leads to a conventionally chosen linearized model\cite{Alouani-Bibi2005,Bell2006,Tzoufras2011}.

Transport equations for high order velocity moments of Boltzmann's equation \cite{Boltzmann1872} or VFP equation\cite{Chen1984} 
can address these challenges, and have demonstrated superior effectiveness in solving problems of plasmas physics.
However, difficulties\cite{Mintzer1965} arise in two aspects: I), the resulting set of equations lack closure because the $l^{th}$-order moment equation contains the moment of order $l+1$. II), the dissipative terms originating from the collision operator are typically nonlinear functions of moments. Consequently, it is necessary to truncate the set of transport equations based on certain assumption about the form of velocity distribution function.
Traditionally, near-equilibrium assumption\cite{Grad1949} is widely adapted in space physics, physics of fluid, plasmas physics and other related fields.

The relaxation process of a system of particles with Coulomb interactions towards a Maxwellian distribution function was initially presented by MacDonald and Rosenbluth\cite{MacDonald1957}. Subsequently, Tanenbaum\cite{Tanenbaum1967} derived the transport equations based on the isotropic Maxwellian distribution function.
The general form of transport equations under near-equilibrium assumption are derived by Chapman\cite{Chapman1916The} and Enskog\cite{Chapman1953}, and extended by Burnett\cite{Burnett1935}.  
Another approach proposed by Grad\cite{Grad1949}, utilizing the Hermite polynomial expansion method, also yields the transport equations. 
Additionally,
Mintzer\cite{Mintzer1965} derive the transport equations by introducing a generalized orthogonal polynomial method, which are capable of describing highly nonequilibrium system. 
These advancements  have been comprehensively reviewed  by Schunk\cite{Schunk1977}. 
However, as highlighted by Schunk\cite{Schunk1977}, both the Chapman-Enskog and Grad procedures exhibit inadequate convergence in highly non-Maxwellian system due to expanding the distribution function into an orthogonal series around a local Maxwellian. These limitations arise from the underlying near-equilibrium assumption. 

Recently, various nonlinear simulations\cite{HuiLi2023,Hao2024Summary} of heat transport in Tokamak plasmas have been invested.
We aim to develop a novel framework\cite{wang2024higherorder} for addressing the nonlinear simulation\cite{Hao2024Summary,Li2024Simulation} for fusion plasmas, under the hypothesis of finitely distinguishable independent\cite{Teicher1963,Yakowitz1968} features (for details, see Sec.\SEC{Finitely distinguishable independent features hypothesis}) rather than relying on the conventional near-equilibrium assumption. This framework is a higher-order moment convergent method, encompassing both a meshfree approach\cite{wang2024Aconservative} and a moment approach. This paper is more directly concerned with introducing the moment approach to derive the transport equations from the 0D-1V VFP equation and novel closure relations for this transport equations.
The Shkarofsky's form of Fokker-Planck-Rosenbluth (FPRS) collision operator\cite{Shkarofsky1963,Shkarofsky1967} is employed to solve the VFP equation,
specifically focusing on the scenario with spherically symmetric velocity space. 
In this situation, we propose a relaxation model based on Maxwellian mixture model (MMM) that effectively captures both near-equilibrium and far-from-equilibrium states.

The remaining sections of this paper are arranged as follows. Sec.\ref{Theoretical formulation} provides an introduction to the VFP equation, RFPS collision operator, and their key properties. In the case of spherical symmetry in velocity space, Sec.\SEC{Relaxation model for homogeneous plasmas} discusses the relaxation model based on MMM. Finally, a summary of our work is presented in Sec.\SEC{Conclusion}.

\end{spacing}

\begin{spacing}{0.7}   

\section{Theoretical formulation}
\label{Theoretical formulation}

\subsection{Vlasov-Fokker-Planck equation}
\label{Vlasov-Fokker-Planck equation}

  The physical state of plasmas is characterized by distribution functions of position vector ${\mathbf{r}}$, velocity vector  ${\mathbf{v}}$ and time $t$, for species $a$, $f=f \left({\mathbf{r}},{\mathbf{v}},t \right)$. 
  In this paper,
  we assume that function $f$ is continuous and exhibits smoothness. The evolution of the system state can be described by the VFP equation
  \cite{Chen1984}. 
  For a homogeneous plasmas system, the VFP equation reduces to the Fokker-Planck collision equation:
  \begin{eqnarray}
      {\frac{\partial}{\partial t}} f \left({\mathbf{v}},t \right)  &=& \mathfrak{C} ~.\label{VFP0D1V}
  \end{eqnarray} 
  The term $\mathfrak{C}$ on the right-hand side of Eq.\EQ{VFP0D1V} represents the Coulomb collision effect on species $a$, encompassing both its self-collision effect of species $a$ and the mutual collision effect between species $a$ and background species (details in Sec.\SEC{Fokker-Planck collision operator}).

The first few moments of the distribution function, such as the mass density $\rho_a(t)$ (zero-order moment), momentum ${\boldsymbol{I}_a}(t)$ (first-order moment), and energy $K_a(t)$ respectively are:
  \begin{eqnarray}
      \rho_a \left(t \right) &=& m_a\left<1,f({\mathbf{v}},t) \right>_{{\mathbf{v}}}, \label{rho}
      \\
      {\boldsymbol{I}_a} \left(t \right) &=& m_a \left<{\mathbf{v}},f({\mathbf{v}},t) \right>_{{\mathbf{v}}}, \label{Ia} 
      \\
      K_a \left(t \right) &=& \frac{m_a}{2} \left<{\mathbf{v}}^2,f({\mathbf{v}},t) \right>_{{\mathbf{v}}} , \label{Ka}
  \end{eqnarray}
where operator $\left<g,h \right>_{{\mathbf{v}}}$ represents the integral of the function $g \cdot h$ with respect to ${\mathbf{v}}$. The temperature at time $t$ is defined as:
  \begin{eqnarray} 
      T_a \left(t \right) &=& \frac{m_a}{3 n_a} \left<({\mathbf{v}} - {\boldsymbol{u}_a})^2,f({\mathbf{v}},t) \right>_{{\mathbf{v}}} . \label{Ta9Ms0D2V}
  \end{eqnarray}
Among them, the average velocity ${\boldsymbol{u}_a} (t) = {\boldsymbol{I}_a} / \rho_a$, number density $n_a (t) = \rho_a / m_a$, momentum amplitude $I_a (t) = \sqrt{{\boldsymbol{I}_a}^2}$. The thermal velocity ${{v}_{ath}} (t) = \sqrt{2 T_a / m_a}$, which depends on $\rho_a, I_a$ and $K_a$, can be expressed as:
\begin{eqnarray}
    {{v}_{ath}} (t) &=& \sqrt{\frac{2}{3} \left(\frac{2 K_a}{\rho_a} - \left(\frac{I_a}{\rho_a} \right)^2 \right)} ~.\label{vath9Ms0D2V}
\end{eqnarray}

\subsection{Fokker-Planck-Rosenbluth collision operator}
\label{Fokker-Planck collision operator}

Without sacrificing generality, the scope of this paper is limited to the case of a two-species plasmas system. In this particular scenario, the collision operator in 
VFP equation represented by Eq.\EQ{VFP0D1V}
will be:
  \begin{eqnarray}
      \mathfrak{C} \left({\mathbf{v}},t \right) &=& {\mathfrak{C}_{ab}} + {\mathfrak{C}_{aa}}  ~,\label{cola}
  \end{eqnarray}
where ${\mathfrak{C}_{ab}}$ and ${\mathfrak{C}_{aa}}$ represent the FPRS\cite{Shkarofsky1963,Shkarofsky1967} collision operator. The mutual collision operator between species $a$ and species $b$ denoted as ${\mathfrak{C}_{ab}}$, is given by:
  \begin{eqnarray}
      {\mathfrak{C}_{ab}} \left({\mathbf{v}},t \right) &=& {\Gamma_{ab}} \left [4 \pi m_M F f + \left(1-m_M \right) {\nabla_{{\mathbf{v}}}} H \cdot {\nabla_{{\mathbf{v}}}} f + 
      \frac{1}{2} {\nabla_{{\mathbf{v}}}} {\nabla_{{\mathbf{v}}}} G : {\nabla_{{\mathbf{v}}}} {\nabla_{{\mathbf{v}}}} f \right]  , \label{FPRS}
  \end{eqnarray}
where ${\Gamma_{ab}}=4\pi \left(\frac{q_a q_b}{4 \pi \varepsilon_0 m_a} \right)^2 {\ln{ \Lambda_{ab}}}$,  $m_M=m_a/m_b$. Here, $m_a$ and $m_b$ represent the mass of species $a$ and $b$ respectively. $q_a$ and $q_b$ denote the charge numbers of species $a$ and $b$. The parameters $\varepsilon_0$ and ${\ln{ \Lambda_{ab}}}$ correspond to the dielectric constant of vacuum and the Coulomb logarithm\cite{Huba2011}.
The function $F=F({\mathbf{v}_b},t)$, representing the distribution function of background species $b$. Functions $H$ and $G$ are the Rosenbluth potentials, which are integral functions of distribution function $F$, reads:
  \begin{eqnarray}
    H({\mathbf{v}},t) &=& \int \frac{1}{\left|{\mathbf{v}}-{\mathbf{v}_b} \right|} F\left({\mathbf{v}_b},t \right) \mathrm{d} {\mathbf{v}} _b, \label{H} \\
    G({\mathbf{v}},t) &=& \int \left|{\mathbf{v}}-{\mathbf{v}_b} \right| F\left({\mathbf{v}_b},t \right) \mathrm{d} {\mathbf{v}} _b ~. \label{G}
  \end{eqnarray} 
By replacing $b$, $F$, and ${\mathbf{v}_b}$ in Eq.\EQ{FPRS} with $a$, $f$, and ${\mathbf{v}}$, respectively, we can derive the FPRS self-collision operator in Eq.\EQ{cola}:
  \begin{eqnarray}
      {\mathfrak{C}_{aa}} \left({\mathbf{v}},t \right) &=& {\Gamma_{aa}} \left (4 \pi f f + \frac{1}{2} {\nabla_{{\mathbf{v}}}} {\nabla_{{\mathbf{v}}}} G : {\nabla_{{\mathbf{v}}}} {\nabla_{{\mathbf{v}}}} f \right)  ~. \label{colaa}
  \end{eqnarray}

The present study exclusively focuses on the scenario where the velocity space exhibits spherical symmetry. When expressing the velocity space in terms of spherical-polar coordinates ${\mathbf{v}}(v, \theta,\phi)$, one can obtain the $(l,m)^{th}$-order amplitude of distribution function by employing a spherical harmonic expansion\cite{Bell2006} (SHE), 
reads:
  \begin{eqnarray}
      f \left({\mathbf{v}},t \right) &=& {\sum_{l=0}^{\infty} } {\sum_{m=-l}^l } {f{_{l}^m}} \left(v ,t \right) {\mathrm{Y}_l^m} \left(\mu, \phi \right) , \label{fhsph}
  \end{eqnarray}
where speed $v = |{\mathbf{v}}|$, $\mu=\cos{\theta}$ and ${\mathrm{Y}_l^m}$ is the spherical harmonic\cite{Arfken1971} without the normalization coefficient, $N_l^m=\sqrt{\frac{2l+1}{4\pi} \frac{(l-m)!}{(l+m)!}}$.
The calculation for the $l^{th}$-order amplitude ${f{_{l}^m}} \left( v,t \right)$ can be obtained through the inverse transformation of Eq.\EQ{fhsph} as follows:
   \begin{eqnarray}
       {f{_{l}^m}} \left({\mathbf{v}},t \right) &=& \frac{1}{(N_l^m)^2} \int_{-1}^1 \int_0^{2 \pi} \frac{1}{2} \left (Y_l^{\left |m \right|} \right)^* f \left ({\mathbf{v}}, t \right) \mathrm{d} \phi  \mathrm{d} \mu, \quad m \ge 0~.  \label{flm0}
   \end{eqnarray}
Therefore, for spherical symmetric velocity space, the amplitude function can be represented as:
\begin{eqnarray}
    {f{_{l}^m}} \left(v,t \right) &=&
    \delta_l^0 \delta_m^0 f({\mathbf{v}},t), \label{flm}
\end{eqnarray}
where $\delta_l^0$ is the Kronecker symbol. 
That implies that the amplitude  function ${f_0^0}$ is non-negative, and higher-order amplitudes ${f{_{l}^m}}$, where $l \ge 1$ 
or $|m| \ge 1$, are all zeros when velocity space is spherically symmetric. 
From now on, we will omit the superscripts of amplitudes due to $m\equiv0$ for scenarios with spherical
symmetric velocity space. For example, we will utilize ${f_0}$ instead of ${f_0^0}$.

We will normalize the velocity-space quantities with the local thermal velocity ${{v}_{ath}}$ for species $a$, ${\hat{\mathbf{v}}}={\mathbf{v}} / {{v}_{ath}}$ and the zeroth-order normalized amplitude function of species $a$ will be:
\begin{eqnarray}
    {\hat{f}_0} \left(\hat{v},t \right) &=&  {\frac{{{v}_{ath}}^3}{n_a}} {f_0} \left(v,t \right)~. \label{fhlm}
\end{eqnarray}
Similarly, the Rosenbluth potentials represented by
Eqs.\EQ{H}-\EQo{G} can be expressed as:
  \begin{eqnarray}
      H(v,t) &=& 4 \pi \frac{n_b}{{{v}_{bth}}} {\hat{H}{_0}},
      \quad
      G(v,t) \ = \ 4 \pi n_b {{v}_{bth}} {\hat{G}{_0}},
  \end{eqnarray}
where the zeroth-order amplitudes of Rosenbluth potentials are:
  \begin{eqnarray}
      {\hat{H}{_0}} \left({{\hat{v}}_{ab}},t \right) &=& \frac{1}{{{\hat{v}}_{ab}}} \left (I_{0,0} + J_{1,0} \right) , \label{Hho0D1V} \\
      {\hat{G}{_0}} \left({{\hat{v}}_{ab}},t \right) &=& {{\hat{v}}_{ab}} \left (\frac{I_{2,0} + J_{1,0}}{3} + I_{0,0} + J_{-1,0} \right)  ~. \label{Gho0D1V}
  \end{eqnarray}
The functions ${I_{i,0}}$ and ${J_{i,0}}$ represent integrals of the normalized background distribution function
${\hat{F}{_0}}\left({{\hat{v}}_b},t \right)$, 
following a similar approach as
Shkarofsky et al.\cite{Shkarofsky1967,Shkarofsky1997} , reads:
  \begin{eqnarray}
      {I_{i,0}} \left({{\hat{v}}_{ab}},t \right) &=& \frac{1}{({{\hat{v}}_{ab}})^i} \int_0^{{{\hat{v}}_{ab}}} {{\hat{v}}_b}^{i+2} {\hat{F}{_0}} \mathrm{d} {{\hat{v}}_b},  \quad i = L,L+2 , \label{IjFo} 
      \\
      {J_{i,0}} \left({{\hat{v}}_{ab}},t \right) &=& ({{\hat{v}}_{ab}})^i \int_{{{\hat{v}}_{ab}}}^{\infty} \frac{{{\hat{v}}_b}^2}{{{\hat{v}}_b}^i} {\hat{F}{_0}} \mathrm{d} {{\hat{v}}_b},  \quad i = L \pm 1  , \ \label{JjFo}
  \end{eqnarray}
where ${{\hat{v}}_{ab}} = {{v}_{abth}} \hat{v}$, ${{v}_{abth}} = {{v}_{ath}} / {{v}_{bth}}$, $\hat{v} = v / {{v}_{ath}}$ and
\begin{eqnarray}
    {\hat{F}{_0}} \left({{\hat{v}}_b},t \right) &=& \frac{{{v}_{bth}}^3}{n_b} F(v_b,t) .
\end{eqnarray}

Therefore, the FPRS collision operator represented by Eq.\EQ{FPRS} can be reformulated as:
  \begin{eqnarray}
      {\mathfrak{C}_{ab}}(v,t) &=& \delta_l^0 {\frac{n_a}{{{v}_{ath}}^3}} {{{\hat{\mathfrak{C}}}{_l}}_{ab}} ~.
  \end{eqnarray}
The $l^{th}$-order normalized amplitude of the mutual FPRS collision operator can be expressed as follows:
  \begin{eqnarray}
      {{{\hat{\mathfrak{C}}}{_l}}_{ab}} \left(\hat{v},t \right) &=& \delta_l^0 4 \pi {\Gamma_{ab}} \left [m_M {\hat{F}{_0}} {\hat{f}_0} + {C_{\hat{H}}} {\frac{\partial {\hat{H}{_0}}}{\partial {\hat{v}_{ab}}}} 
      {\frac{\mathrm{\partial} {\hat{f}_0}}{\mathrm{\partial} \hat{{v}}}} + 2 \frac{{C_{\hat{G}}}}{{{\hat{v}}_{ab}} \hat{v}}  {\frac{\partial {\hat{G}{_0}}}{\partial {\hat{v}_{ab}}}} {\frac{\mathrm{\partial} {\hat{f}_0}}{\mathrm{\partial} \hat{{v}}}} + 
      {C_{\hat{G}}} {\frac{\partial^2 {\hat{G}{_0}}}{\partial {{\hat{v}_{ab}}}^2}}  {\frac{\mathrm{\partial}^2 {\hat{f}_0}}{\mathrm{\partial} {\hat{{v}}}^2}} \right] 
      . \label{FPRS0D1V}
  \end{eqnarray}
The coefficients
  \begin{eqnarray}
      {C_{\hat{H}}}= \frac{1-m_M}{{{v}_{abth}}} , \quad 
      {C_{\hat{G}}}= \frac{1}{2 \left({{v}_{abth}} \right)^2}
      ~. \label{CFHGh}          
  \end{eqnarray}

  The FPRS collision operator represented by Eq.\EQ{FPRS0D1V} is a special instance of the scenario with axisymmetric velocity space as described in Ref.\cite{wang2024Aconservative}.
  Similarly, the $l^{th}$-order normalized amplitude of the self-collision operator can be expressed as:
  \begin{eqnarray}
      {{{\hat{\mathfrak{C}}}{_l}}_{aa}} \left(\hat{v},t \right) &=& \delta_l^0 4 \pi {\Gamma_{aa}} \left ({\hat{f}_0} {\hat{f}_0} + \frac{1}{ \hat{v}^2}  {\frac{\partial {\hat{G}{_0}}}{\partial {\hat{v}}}} {\frac{\mathrm{\partial} {\hat{f}_0}}{\mathrm{\partial} \hat{{v}}}} + 
      \frac{1}{2} {\frac{\partial^2 {\hat{G}{_0}}}{\partial {{\hat{v}}}^2}} {\frac{\mathrm{\partial}^2 {\hat{f}_0}}{\mathrm{\partial} {\hat{{v}}}^2}} \right) ~. \label{colaa0D1V}
  \end{eqnarray}
  Applying Eq.\EQ{cola}, the $l^{th}$-order normalized amplitude of FPRS collision operator will be:
  \begin{eqnarray}
      {{{\hat{\mathfrak{C}}}{_l}}} \left(v,t \right) &=& {\frac{n_b}{{{v}_{bth}}^3}} {{{\hat{\mathfrak{C}}}{_l}}_{ab}} + {\frac{n_a}{{{v}_{ath}}^3}} {{{\hat{\mathfrak{C}}}{_l}}_{aa}} \label{colhla}.
  \end{eqnarray}
  Therefore, when velocity space exhibits spherical symmetry, the VFP equation represented by Eq.\EQ{VFP0D1V} can be rewritten as:
  \begin{eqnarray}
      {\frac{\partial}{\partial t}} {f_l} \left(v,t \right) &=& \delta_l^0 {\frac{n_a}{{{v}_{ath}}^3}} {{{\hat{\mathfrak{C}}}{_l}}}  ~.\label{VFPl}
  \end{eqnarray}
 The equation mentioned above will be referred to as 0D-1V VFP spectrum equation.

\subsection{Elementary properties of FPRS collision operator} 
\label{Elementary properties of FPRS collision operator}

Firstly, we give the definitions of $(j,l)^{th}$-order kinetic moment:
\begin{eqnarray}
    {\mathcal{M}_{j,l}} \left(t \right)  &=&  4 \pi \rho_a ({{v}_{ath}})^j \int_0^{\infty} \hat{v}^{j+2} {\hat{f}_l} \mathrm{d} \hat{v} , \quad j \ge -2-l . \label{Mjl3D2V}
\end{eqnarray}
Specially, the mass density\EQgive{rho} and energy\EQgive{Ka} can be expressed as:
  \begin{eqnarray}
      \rho_a \left(t \right) &=& \mathcal{M}_{0,0} ,  \label{naM} 
      \quad
      K_a \left(t \right) \ = \ \frac{1}{2} \mathcal{M}_{2,0} \label{KaM}
  \end{eqnarray}
 and the momentum will always be zero in spherically symmetric velocity space. Therefore, the thermal velocity\EQ{vath9Ms0D2V} can be rewritten as:
\begin{eqnarray}
    {{v}_{ath}} (t) &=& \sqrt{\frac{2}{3} \frac{\mathcal{M}_{2,0}}{\mathcal{M}_{0,0}} } ~.\label{vathM}
\end{eqnarray}
 
Similar to Eq.\EQ{Mjl3D2V}, the $(j,l)^{th}$-order kinetic dissipative force is defined as:
\begin{eqnarray}
    {\mathcal{R}_{j,l}} \left(t \right)  &=&  4 \pi \rho_a ({{v}_{ath}})^j
    \int_0^{\infty} \hat{v}^{j+2} {{{\hat{\mathfrak{C}}}{_l}}} \mathrm{d} \hat{v} , \quad j \ge -2-l ~. \label{Rjl3D2V}
\end{eqnarray}
Please note that ${{\mathcal{R}}{_{0,0}}} \equiv 0$ for all elastic collisions and $\mathcal{R}_{j,l \ge 1} \equiv 0$ 
in the scenario of spherically symmetric
velocity space.

The FPRS collision operator theoretically ensures the conservation of mass, momentum, and energy during the collision process between two species. When the velocity space exhibits spherical symmetry, it can be expressed as follows: 
  \begin{eqnarray}
      {{{\mathcal{R}}{_{ab}}}{_{0,0}}} &=& {{{\mathcal{R}}{_{ba}}}{_{0,0}}} \ = \ 0, \label{Cnh0D1V}
      \\
      \frac{1}{3} {{{\mathcal{R}}{_{ab}}}{_{1,1}}} &=& - \frac{1}{3} {{{\mathcal{R}}{_{ba}}}{_{1,1}}} \ = \ 0, \label{CIh0D1V}
      \\
      \frac{1}{2} {\mathcal{R}}{_{ab}}_{2,0} &=& - \frac{1}{2} {\mathcal{R}}{_{ba}}_{2,0}~. \label{CKh0D1V}
  \end{eqnarray}
  Here, function ${{\mathcal{R}}{_{ab}}_{j,l}^{0}}$ represents the $(j,l)^{th}$-order kinetic dissipative force exerted on species $a$ during mutual collisions with species $b$.

\section{Relaxation model for homogeneous plasmas}
\label{Relaxation model for homogeneous plasmas}

The starting point for the derivation of transport equations for plasmas is VFP equation\EQ{VFP0D1V}. These equations can be obtained by multiplying the both side of VFP equation by an appropriate function of velocity $g= g({\mathbf{v}})$ and then integrating over all velocity space. 

\subsection{Transport equations}
\label{Transport equations}

In the spherical coordinate system, by multiplying both sides of Eq.\EQ{VFPl} by $4 \pi m_a v^{j+2} \mathrm{d} v$ and integrating over the semi-infinite interval $v = \left [0, \infty \right)$, and then applying Eqs.\EQ{Mjl3D2V}-\EQo{Rjl3D2V}, we obtain the $(j,l)^{th}$-order transport equation (or kinetic moment evolution equation) as follows:
\begin{eqnarray}
    {\frac{\partial}{\partial t}} {\mathcal{M}_{j,l}} \left(t \right) &=& \delta_l^0  \rho_a \left({{v}_{ath}} \right)^{j} {{\hat{\mathcal{R}}}{_{j,0}}}, \ j \ge -2-l , \label{dtMho}
\end{eqnarray}
where the normalized kinetic dissipative force, $ {\hat{\mathcal{R}}_{j,l}} = {\mathcal{R}_{j,l}} / [\rho_a ({{v}_{ath}})^j]$, is given by
\begin{eqnarray}
    {{\hat{\mathcal{R}}}{_{j,0}}} \left(t \right) &=& {\frac{n_b}{{{v}_{bth}}^3}} {\Gamma_{ab}} {{{\hat{\mathcal{R}}}{_{ab}}}{_{j,0}}}  + {\frac{n_a}{{{v}_{ath}}^3}} {\Gamma_{aa}} {{{\hat{\mathcal{R}}}{_{aa}}}{_{j,0}}} 
      ~.\label{Rhj0a0D1V}
\end{eqnarray}
Regard Eq.\EQ{Rhj0a0D1V} as the kinetic dissipative force closure relation,
representing the inherent nonlinear relation between kinetic moments and kinetic dissipative forces.
The first few orders of transport equations\EQ{dtMho} associated with conserved moments can be expressed as:
\begin{eqnarray}
    {\frac{\partial}{\partial t}} \rho_a \left(t \right) &=& \rho_a {{\hat{\mathcal{R}}}{_{0,0}}}, \label{dtna}
    \\
    {\frac{\partial}{\partial t}} I_a \left(t \right) &=& \frac{1}{3} \rho_a {{v}_{ath}} {{\hat{\mathcal{R}}}{_{1,1}}}, \label{dtIa}
    \\
    {\frac{\partial}{\partial t}} K_a \left(t \right) &=& \frac{1}{2} \rho_a \left({{v}_{ath}} \right)^{2} \hat{\mathcal{R}}_{2,0} ~.\label{dtKa}
\end{eqnarray} 
The transport equation\EQ{dtMho} can be solved by a Runge-Kutta solver, such as trapezoidal\cite{Rackauckas2017} scheme,
similar to the meshfree approach in Re\cite{wang2024Aconservative}.

\subsection{Finitely distinguishable independent features hypothesis}
\label{Finitely distinguishable independent features hypothesis}

Boltzmann \cite{Boltzmann1872} proved that in a thermodynamic equilibrium, the velocity space  exhibits spherical symmetry and the distribution function follows a Maxwellian distribution, 
      \begin{eqnarray}
            f \left({\mathbf{v}},t \right)  &=& \frac{1}{\pi^{3/2}} \frac{n_a}{({{v}_{ath}})^3} \exp{\left[-\frac{{\mathbf{v}}^2}{\left({{v}_{ath}} \right)^2} \right]} ~.  \label{fM}
      \end{eqnarray}
According to  Eq.\EQ{flm} and Eq.\EQ{fhlm}, the above equation can be expressed in normalized form in a spherical-polar coordinate system as:
      \begin{eqnarray}
            {\hat{f}_l} \left(\hat{v},t \right)  &=& \delta_l^0 \frac{1}{\pi^{3/2}} e^{- \hat{v}^2} ~.  \label{fho}
      \end{eqnarray}
Let Eq.\EQ{fho} represent the Maxwellian model (MM).  

In the more general case, the velocity space of the system exhibits spherical symmetry but may not be in a state of thermodynamic equilibrium. Under this circumstance,  the one-dimensional amplitude function ${\hat{f}_0}$ can be approximated by a linear combination of King functions ${\mathcal{K}_0}$, 
named as King function expansion (KFE), reads:
    \begin{eqnarray}
        {\hat{f}_l} \left(\hat{v},t \right)  &=& \delta_l^0 \frac{\sqrt{2 \pi}}{\pi^{3/2}} \sum_{r=1}^{{N_{K_a}}} {\hat{n}_{a_r}} {\mathcal{K}_0} \left(\hat{v};{\hat{u}_{a_r}},{{\hat{v}}_{{ath}_r}} \right)
        ,  \label{fhoKo}
    \end{eqnarray}
where ${N_{K_a}} \in \mathbb{N}^+$.  
The parameters, ${\hat{n}_{a_r}}={n_{a_r}} / n_a$, ${\hat{u}_{a_r}}={u_{a_r}} / {{v}_{ath}}$ and ${{\hat{v}}_{{ath}_r}} = {{v}_{{ath}_r}} / {{v}_{ath}}$, 
are the characteristic parameters of $r^{th}$ sub-distribution of ${\hat{f}_0}$.
The King function is defined as follows:
    \begin{eqnarray}
        {\mathcal{K}_0} \left(\hat{v};\iota, \sigma \right) &=& \frac{1}{\sqrt{2 \pi}} \frac{1}{\sigma^3 } \frac{\sigma^2}{2 \iota \hat{v} } \exp{\left(-\frac{\hat{v}^2 + \iota^2}{\sigma^2} \right)} \sinh{\left(\frac{2 \iota \hat{v} }{\sigma^2} \right)} ~.  \label{K0}
    \end{eqnarray}
Let Eq.\EQ{fhoKo} represent the zeroth-order King mixture model (KMM0), indicating that the plasmas are in a quasi-equilibrium state.
Similarly, the normalized amplitudes of background distribution function
can be approximated as:
    \begin{eqnarray}
        {\hat{F}_L} \left({\hat{v}}_b,t \right)  &=&  \delta_L^0 \frac{\sqrt{2 \pi}}{\pi^{3/2}} \sum_{s=1}^{{N_{K_b}}} {\hat{n}_{b_s}} {\mathcal{K}_0} \left(\hat{v};{\hat{u}_{b_s}},{{\hat{v}}_{{bth}_s}} \right)
        ~.  \label{FhoFhos}
    \end{eqnarray}
    
If two known groups of characteristic parameters, $\left (\iota_1, \sigma_1 \right) $ and $\left (\iota_2, \sigma_2 \right) $, each with respective weights
${\hat{n}_{a_1}}$ and ${\hat{n}_{a_2}}$, satisfy
    \begin{eqnarray}
      \left| \frac{\sigma_1}{\sigma_2} - 1 \right| + \left| \frac{\iota_1}{\iota_2} - 1 \right|  & \le & rtol, \label{Dvhths}
    \end{eqnarray}
we claim that the King function ${\mathcal{K}_l} \left(v;\iota_1,\sigma_1 \right)$ and ${\mathcal{K}_l} \left(v;\iota_2,\sigma_2 \right)$ are identical with parameters $(\iota_0,\sigma_0)$.
Here, $rtol$ is a given relative tolerance with a default value, $rtol=10^{-10}$. The weight of ${\mathcal{K}_l} \left(v;\iota_0,\sigma_0 \right)$
is given by ${\hat{n}_{a_0}} = {\hat{n}_{a_1}} + {\hat{n}_{a_2}}$. Eq.\EQ{Dvhths} serves as the indistinguishable condition of the King function.

  \begin{figure}[htbp]
	\begin{center}
		\includegraphics[width=0.7\linewidth]{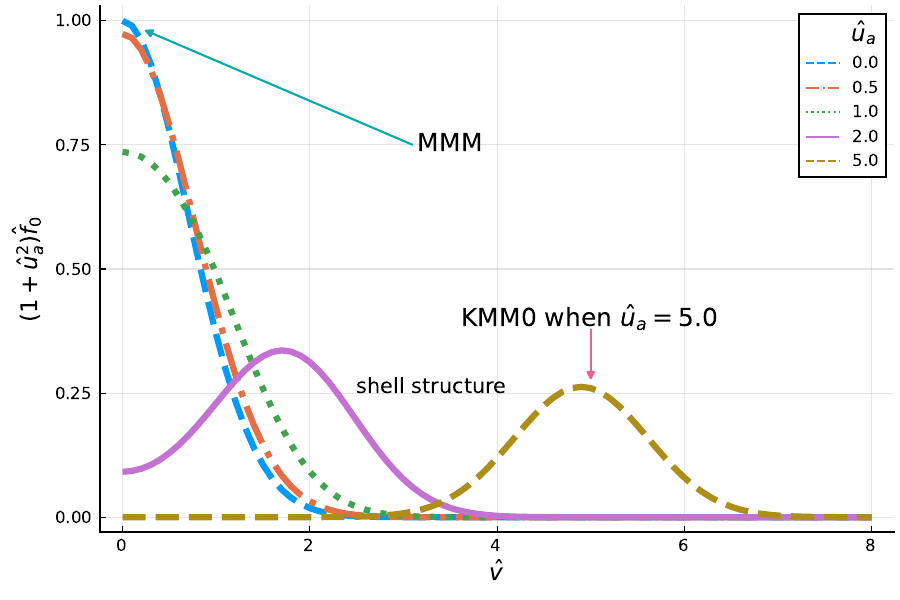}
	\end{center}
	\caption{Illustration of the velocity distribution functions multiplied by a factor $(1+\hat{u}_a^2)$ for ${N_{K_a}} \equiv 1$ and various normalized average velocity $\hat{u}_a$.}
	\label{FigKMM0}
  \end{figure}

The above model is under the finitely distinguishable independent\cite{Teicher1963,Yakowitz1968} features (FDIF) hypothesis. This hypothesis posits that given 
indistinguishable condition\EQ{Dvhths}, a finite-volume, finite-density, finite-temperature, and finite-component fully ionized plasmas system has a finite number of distinguishable independent characteristics.
This hypothesis indicates that $N_{k_a}$ is a finite-size number in KMM0\EQ{fhoKo}.

The velocity shell structure\cite{Min2015} is a typical characteristic feature for $\alpha$ particle distribution function\cite{Gorelenkov2014} in burning plasmas. 
When $\sum_r(|{\hat{u}_{a_r}}|^2)$ in KMM0\EQ{fhoKo} is greater than zero, we call that the distribution function described by KMM0 has velocity shell structure.
This structure can be observed in Fig.\FIG{FigKMM0}, particularly when $\hat{u}_a > 1$.

When there is no shell structure in velocity space for the distribution function, we can simplify Eq.\EQ{fhoKo} using $\sum_r(|{\hat{u}_{a_r}}|^2) \equiv 0$ as 
    \begin{eqnarray}
        {\hat{f}_l} \left(\hat{v},t \right)  &=& \delta_l^0 \frac{1}{\pi^{3/2}} \sum_{r=1}^{{N_{K_a}}} \left[\frac{{\hat{n}_{a_r}}}{{{\hat{v}}_{{ath}_r}}^3} \exp{\left(-\frac{\hat{v}^2}{{{\hat{v}}_{{ath}_r}}^2} \right)} \right]
        ~.  \label{fhofhor}
    \end{eqnarray}
The Maxwellian mixture model (MMM), denoted by Eq.\EQ{fhofhor}, represents a shell-less distribution, which indicates that the plasmas are in a shell-less quasi-equilibrium state.
Fig.\FIG{FigKMM0} illustrates velocity distribution functions described by KMM0 (including MMM) as a function of $\hat{v}$, along with various normalized average velocity $\hat{u}_a$ when ${N_{K_a}} \equiv 1$. To examine the details of cases where $\hat{u}_a > 1$, the distribution function is multiplied by a factor of $(1+\hat{u}_a^2)$.
The convergence of KMM0 and MMM can be proved based on 
Wiener's Tauberian theorem\cite{Wiener1932,Vladimirov1988,Korevaar2004}. 
The proof is provided in Appendix\APP{Convergence of KMM0 and MMM}.

The various advantages and disadvantages of KFE are outlined in a topic review\cite{wang2024higherorder} of NLVFP
code for solving the 0D-2V nonlinear VFP equation. Here, we represent the ones for scenarios with
spherical symmetric velocity space, comparing to the conventional methods, such as finite difference
method\cite{Tzoufras2011} (FDM), particle-in-cell\cite{Thomas2012,Pfeiffer2017} (PIC) method,  Laguerre polynomial expansion (LPE) method
and Hermite polynomial expansion\cite{Li2021} (HPE) method.
\subsubsection{Some advantages of KFE}
\label{Some advantages of KFE}

\begin{spacing}{0.2}  
\begin{itemize}
    \item[(i)] Coulomb collisions lead to rapid convergence for the amplitudes in speed coordinate. As demonstrated in Ref.\cite{wang2024Aconservative}, the KFE is a moment convergent technique that has been demonstrated to
    achieve up to order 16, making it significantly faster than FDM which typically achieves convergence order of no more than 5.
    \item[(vi)] The King function, serving as a one-dimensional continuous smooth function, ensures that KFE
    produces results without noise over a PIC method and more stable than LPE method. Additionally, achieving higher-order moment convergence remains a challenge for the PIC method.
    
    \item[(vii)] KFE effectively captures the complete nonlinear effects of the VFP equation by employing adaptive values of ${N_{K_a}}$ and ${N_{K_b}}$, typically no more than 20 for a weakly anisotropic plasmas\cite{wang2024Aconservative}, due
    to the rapid convergence of KFE. In contrast, HPE methods typically depict effects where not
    far from the thermodynamic equilibrium state, which is usually based on the near-equilibrium
    assumption.
    \item[(viii)] KFE ensures the symmetry of the collision operator in discrete\cite{wang2024Aconservative} due to its rapid convergence.
    \item[(iv)] KFE has the capability to naturally capture the isotropic Maxwellian state in velocity space.
    Additionally, the resulting solution exhibits robustness comparing to FDM/PIC/LPE methods.
\end{itemize}
\end{spacing}

\subsubsection{Some disadvantages of KFE}
\label{Some disadvantages of KFE}

\begin{itemize}
    \item[(i)] The presence of full nonlinearity in speed coordinate results in a nonlinear algebraic dependencies
    between the desired moments and kinetic dissipative forces.
    \item[(ii)] The characteristic parameters in KFE should be determined using alternative methods\cite{wang2024Aconservative}, such
    as solving the characteristic parameter equations as described bellow.
    \item[(iii)] The entire solution process depends on numerical solutions and appears complex.
\end{itemize}
\noindent
The above challenges are inherent outcomes of the nonlinearity and have not been observed to pose
significant issues individually, as evidenced in a more complex scenario involving the solution of 0D-2V
VFP equation\cite{wang2024Aconservative}. 
However, the benefits is obvious that KFE can effectively handle scenarios where
near-equilibrium assumption fails.

Expansion in King function might seem complicated, but in fact the equations turn out to be
fairly straightforward and robust due to the following two reasons:
\begin{itemize}
    \item[(A)] Many natural statistical systems, including plasmas system, are independent and identically distributed systems that often adhere to the central limit theorem\cite{Rosenblatt1956}, allowing their probability density functions to be described utilizing Gaussian functions.
    \item[(B)] Gaussian function is simple and King function denoted by Eq.\EQ{fhoKo} is the zeroth-order amplitude of the Gaussian function in spherical-polar coordinate system, which can be obtained by
    employing SHE.
\end{itemize} 

It is a well-established  fact that plasmas problems can be described by various reduced models\cite{Thomas2012},
including the traditional VFP approach\cite{Taitano2016}, particle approach\cite{xiao2024bohm} and traditional moments approach such
as Grad's moment method\cite{Grad1949}. A direct numerical code can effectively address the VFP equation\EQ{VFP0D1V} or
VFP spectrum equation\EQ{VFPl} for isotropic velocity distribution of homogeneous plasmas when nonlinearity is not strong or important, and may be more effective than KFE approach. However, plasmas are
typically a complex system at multi-time and multi-space scales. Phenomena such as turbulent transport\cite{HuiLi2023,Hao2024Summary} and $\alpha$ particles heating\cite{Sharapov2013} general are nonlinear and affected by the higher-order moments.
Preserving the inherent nonlinearity and ensuring the higher-order moments convergence with conservation laws simultaneously are crucial for simulating the evolution of these plasmas systems. This poses
a challenge for traditional approaches, but can effectively addressed by the KFE approach under the
FDIF hypothesis, particularly in multi-dimensional and multi-velocity plasmas systems\cite{wang2024Aconservative}.

\subsection{Characteristic parameter equation}
\label{Characteristic parameter equation}

Determining the unknown characteristic parameters in KFE\EQ{fhoKo} based on the values of ${\hat{f}_l}$ in the
meshfree approach\cite{wang2024Aconservative} or directly using kinetic moments ${\mathcal{M}_{j,l}}$ are both intermediate parameters methods.
Under FDIF hypothesis, we utilize a parametric equation approach to characterize the intrinsic nonlinear correlation between kinetic moments and kinetic dissipative forces represented by Eq.\EQ{Rhj0a0D1V}. The
intermediate parameters can be obtained by solving the following characteristic parameter equations
(CPEs).
Substituting Eq.\EQ{fhoKo} into the definition of kinetic moment\EQ{Mjl3D2V}, and simplifying the result yields the CPEs when velocity space of the system exhibits spherical symmetry, namely:
      \begin{eqnarray}
            {\mathcal{M}_{j,l}} \left(t \right) = \delta_l^0 {C_M}_j^0 \rho_a \left({{v}_{ath}} \right)^j \sum_{r=1}^{{N_{K_a}}} {\hat{n}_{a_r}} \left({{\hat{v}}_{{ath}_r}} \right)^j 
            \left [1 + \sum_{\beta=1}^{j/2} C_{j,0}^\beta \left (\frac{{\hat{u}_{a_r}}}{{{\hat{v}}_{{ath}_r}}} \right)^{2 \beta} \right], j \in \left\{(2j_p-2)|j_p \in \mathbb{N}^+ \right\} ~. \label{CPEs9Ms0D2V}
      \end{eqnarray}
The coefficient 
      \begin{eqnarray}
          {C_M}_j^0  &=&  \frac{(j + 1)!!}{2^{j/2}} ,   \label{CMj} 
          \quad
          C_{j,0}^\beta \ = \ 2^\beta \frac{C_\beta^{j/2}}{(2 \beta + 1)!!},
      \end{eqnarray}
where $C_\beta^{j/2}$ is the binomial coefficient.
Similarly, Substituting Eq.\EQ{fhofhor} into 
the definitions of kinetic
moment represented by
Eq.\EQ{Mjl3D2V} gives:
      \begin{eqnarray}
            {\mathcal{M}_{j,l}} \left(t \right) &=& \delta_l^0 {C_M}_j^0 \rho_a \left({{v}_{ath}} \right)^j \sum_{r=1}^{{N_{K_a}}} {\hat{n}_{a_r}} \left({{\hat{v}}_{{ath}_r}} \right)^j, \quad j \in \left\{(2j_p-2)|j_p \in \mathbb{N}^+ \right\} ~. \label{CPEs9Ms0D1V}
      \end{eqnarray}
In particular, when $j=2$, we obtain:
  \begin{eqnarray}
      \mathcal{M}_{2,0} \left(t \right) &=&  2 K_a ~. \label{KaMh}
  \end{eqnarray}

Generally, the CPEs\EQ{CPEs9Ms0D2V} typically are a set of nonlinear algebraic equations, encompassing a total of $3 {N_{K_a}}$ unidentified parameters or $2 {N_{K_a}}$ unidentified parameters in Eq.\EQ{CPEs9Ms0D1V}. If we have knowledge of $3 N_{k_a}$ kinetic moments ${{\mathcal{M}}{_{j,0}}}$, solving the well-posed CPEs can provide us with all the characteristic parameters in Eq.\EQ{fhoKo} or Eq.\EQ{fhofhor}. 
The updated values of ${{\mathcal{M}}{_{j,0}}}$ can be determined by solving the transport equations\EQ{dtMho} utilizing Runge-Kutta method. However, there will be discrete errors in time due to a finite timestep, and Eq.\EQ{CPEs9Ms0D2V} will not be exactly satisfied. To show this, we consider an implicit Euler scheme for simplicity, which gives
    \begin{eqnarray}
      {{\mathcal{M}}{_{j,0}}}({t_{k+1}}) &=& {{\mathcal{M}}{_{j,0}}}({t_k}) +{{\Delta}_{{t_k}}} {\frac{\partial}{\partial t}} {{\mathcal{M}}{_{j,0}}}({t_{k+1}}) + {\mathscr{O}} ((\Delta {t_k})^p),  \label{Mjltk1}
    \end{eqnarray}
where ${t_k}$ represents the $k^{th}$-level of time step and $\Delta {t_k}={t_{k+1}}-{t_k}$, denoting the current time-step size. Since Eq.\EQ{CPEs9Ms0D2V} represents a set of nonlinear algebraic equations, the order of time-discrete errors, $p$, must be no less than 2 in an Euler scheme according to Taylor expansion. While a higher-order time integration scheme can mitigate the time-discrete errors, it cannot completely eliminate them. This paper specifically concentrates on error reduction in the velocity space of the VFP equation using a moment approach. A future study will investigate the convergence of time-discrete order within this framework.

An optimization technique utilizing least squares method\cite{Fong2011} (LSM) is presented in Ref.\cite{wang2024Aconservative} for solving the CPEs with a specified collection of $j$, denoted as $[j]$ and the updated values of the kinetic moments, denoted as $M_{j,l}({t_{k+1}})$. The relative deviation between $M_{j,l}({t_{k+1}})$ and the target kinetic moment at $(k+1)^{th}$ time level, ${{\mathcal{M}}{_{j,0}}}({t_{k+1}})$, is referred to:
    \begin{eqnarray}
      \delta {{\mathcal{M}}{_{j,0}}}({t_{k+1}}) &=& \left| \frac{{{\mathcal{M}}{_{j,0}}}({t_{k+1}}) - M_{j,0}({t_{k+1}})}{M_{j,0}({t_{k+1}})} \right| .  \label{RDMjltk1}
    \end{eqnarray}
Notes that the value of $M_{j,0}({t_{k+1}})$ is numerically approximated with time discrete errors utilizing Eq.\EQ{Mjltk1}, while ${{\mathcal{M}}{_{j,0}}}({t_{k+1}})$ exactly satisfies the form of Eq.\EQ{CPEs9Ms0D1V}, representing the desired kinetic moment at $(k+1)^{th}$ time level under the FDIF hypothesis. The collection $[j]$ are not unique, which can be composed by any order in theoretically. Two schemes for $[j]$ are provided in Ref.\cite{wang2024Aconservative}, and a general scheme will be published in our future work.

Additionally, the adaptivity of ${N_{K_a}}$ is crucial for capturing nonlinear effects, and it should be noted that the scheme is also not exclusive. Here, we present a brief overview of the approach utilized in Ref.\cite{wang2024Aconservative}, which has been demonstrated to be effective for weakly anisotropic plasma. 
If $\sum_j \delta {{\mathcal{M}}{_{j,0}}}({t_{k+1}}) \ge Rtol, j \in [j]$ at the initial stage of $(k+1)^{th}$ time level in an implicit Euler scheme, the number of King functions at the $(k+1)^{th}$ time level, ${N_{K_a}}({t_{k+1}})$, will increase by one until reaching a specified maximum value, ${N_K^{max}}$. Here, $Rtol$ is a predefined relative tolerance, for example, $Rtol=10^{-6}$.
Then ${N_{K_a}}({t_{k+1}})$ will be determined at the following stage of $(k+1)^{th}$ time level by utilizing the following strategy:
    \begin{eqnarray}
      {N_{K_a}}({t_{k+1}}) &=& {N_{K_a}}({t_k}) - d{N_{K_a}}({t_k}),  \label{NKa}
      \\
      d{N_{K_a}}({t_k}) &=&
        \begin{aligned}
            \left \{
            \begin{array}{cc}
                -1, & Eq.\EQ{Dvhths} == false,  \\
                1, & Eq.\EQ{Dvhths} == true  ~.
            \end{array}
            \right .
        \end{aligned}
         \label{dNK}
    \end{eqnarray}

The key to utilizing an adaptive ${N_{K_a}}$ lies in the characteristic parameters that encompass all information for kinetic moments of any order under the FDIF hypothesis. Therefore, when ${N_{K_a}}({t_{k+1}}) > {N_{K_a}}({t_k})$, we can include new independent kinetic moments to establish a new well-posed CPEs, obtaining ${{\mathcal{M}}{_{j,0}}}({t_k})$ according to Eq.\EQ{CPEs9Ms0D2V} and ${{\hat{\mathcal{R}}}{_{j,0}}}({t_{k+1}})$ according to Eq.\EQ{Rhabj000D1V}. 
When ${N_{K_a}}({t_{k+1}}) < {N_{K_a}}({t_k})$, we can eliminate higher-order transport equations, which are typically more intricate, to obtain a new effective well-posed CPEs.
The procedures for determining ${N_{K_a}}({t_{k+1}})$ and optimizing the target kinetic moment ${{\mathcal{M}}{_{j,0}}}({t_{k+1}})$ are outlined in Algorithm \ref{alg: NKa}.

    \begin{algorithm}
    \caption{
    Algorithm for determining ${N_{K_a}}({t_{k+1}})$ and optimizing the target kinetic moment ${{\mathcal{M}}{_{j,0}}}({t_{k+1}})$ at the $(k+1)^{th}$ time level.} \label{alg: NKa}
    
      Update ${\frac{\partial}{\partial t}} {{\mathcal{M}}{_{j,0}}}({t_{k+1}})$ according to Eq.\EQ{dtMho}

      \textbf{For} $\gamma = 1,2,\cdots, N_{in}$ \footnote{Notes: The symbol "$N_{in}$" represents the maximum number of iterations in the implicit Euler scheme.}

      Calculate $M_{j,0}({t_{k+1}})$ according to Eq.\EQ{Mjltk1}, update $n_a({t_{k+1}})$ and ${{v}_{ath}}({t_{k+1}})$
      
      \textbf{If} $\gamma = 1$
      
           \quad ${N_{K_a}}({t_{k+1}}) = {N_{K_a}}({t_k}) + 1$
      
           \quad Let ${{\mathcal{M}}{_{j,0}}}({t_{k+1}})=M_{j,0}({t_{k+1}})$
           
           \quad Compute the new characteristic parameters by solving CPEs\EQ{CPEs9Ms0D2V}

          \quad \textbf{If} $\delta {{\mathcal{M}}{_{j,0}}}({t_{k+1}}) \le Rtol$ 

           \quad \quad \textbf{Break}
          
          \quad \textbf{End}
           
      \textbf{Else}

         \quad Compute $d{N_{K_a}}({t_k})$ and update ${N_{K_a}}({t_k})$
      
           \quad Let ${{\mathcal{M}}{_{j,0}}}({t_{k+1}}) = M_{j,0}({t_{k+1}})$, updated CPEs by adjusting the collection $[j]$
           
           \quad Compute the new characteristic parameters by solving the updated well-posed CPEs\EQ{CPEs9Ms0D2V}
          
          \quad \textbf{If} $\delta {{\mathcal{M}}{_{j,0}}}({t_{k+1}}) \le Rtol$ 

           \quad \quad \textbf{Break}
          
          \quad \textbf{End}
          
          \quad \textbf{If} ${N_{K_a}} \ge {N_K^{max}}$ 

           \quad \quad \textbf{Warning}: Checking the effective of timestep ${{\Delta}_{{t_k}}}$, or parameters (${N_K^{max}}$, $Rtol$)

           \quad \quad Reduce the timestep, ${{\Delta}_{{t_k}}} = {{\Delta}_{{t_k}}}/2$, go back to step 1
          
          \quad  \textbf{End}

      \textbf{End}
      
      \textbf{End}

    \end{algorithm}

\subsection{Kinetic moment-closed model based on MMM}
\label{Kinetic moment-closed model based on MMM}

The analytical expression of the $(j,l)^{th}$-order normalized kinetic dissipative force\EQ{Rhj0a0D1V} can be obtained by substituting Eqs.\EQ{fhoKo}-\EQo{fhofhor} and\EQ{FhoFhos} into Eq.\EQ{FPRS0D1V}, and then applying Eq.\EQ{Rjl3D2V}. When the velocity space exhibits spherical symmetry without shell structure, which means $\sum_r(|{\hat{u}_{a_r}}|^2)  \equiv 0$  and $\sum_s(|{\hat{u}_{b_s}}|^2)  \equiv 0$, this expression will be:
\begin{eqnarray}
  \begin{aligned}
    {{\hat{\mathcal{R}}}{_{ab}}_{j,l}} \left(t \right) \ =& \ \delta_l^0 \frac{8}{\pi} {\Gamma} \left(\frac{3+j}{2} \right) \sum_{s=1}^{{N_{K_b}}} {\hat{n}_{b_s}}  \sum_{r=1}^{{N_{K_a}}} {\hat{n}_{a_r}} \frac{1}{{{\hat{v}}_{{ath}_r}}^7}  \left [ m_M {{\hat{v}}_{{ath}_r}} \left(\frac{1 + {{v}_{abth}}^2}{{{\hat{v}}_{{ath}_r}}^2} \right)^{-(3+j)/2} - 
      \right. \\& \left.
    \left({{\hat{v}}_{{ath}_r}} \right)^{3+j} \frac{{{\hat{v}}_{{bth}_s}}^2-m_M {{v}_{abth}}^2 {{\hat{v}}_{{ath}_r}}^2}{{{v}_{abth}}^4 {{\hat{v}}_{{bth}_s}}} {\mathrm{_2 F_1}} \left(\frac{1}{2},\frac{3+j}{2},\frac{3}{2},-\frac{{{v}_{abth}}^2 {{\hat{v}}_{{ath}_r}}^2}{{{\hat{v}}_{{bth}_s}}^2}\right) +  \ 
    \right. \\& \left.
    \left (m_M - 1 - \frac{{{\hat{v}}_{{bth}_s}}}{{{v}_{abth}}^4} \right) \frac{{{\hat{v}}_{{ath}_r}}^2}{{{v}_{abth}}^2 {{\hat{v}}_{{bth}_s}}} \left(\frac{1}{{{\hat{v}}_{{ath}_r}}^2} + \frac{{{v}_{abth}}^2}{{{\hat{v}}_{{bth}_s}}^2}\right)^{-(3+j)/2}
      \right],  \quad j \ge -2
      , \label{Rhabj000D1V}
  \end{aligned}
\end{eqnarray}
where ${\mathrm{_2 F_1}}(a,b,c,z)$ represents the Gauss hypergeometric2F1\cite{Arfken1971} function of the variable $z$.
Similarly, in a self-collision process with $m_M \equiv 1$ and ${{v}_{abth}}\equiv 1$, the normalized dissipative force can be obtained as presented:
\begin{eqnarray}
  \begin{aligned}
    {{\hat{\mathcal{R}}}{_{aa}}_{j,l}} \left(t \right) \ =& \ \delta_l^0 \frac{8}{\pi} {\Gamma} \left(\frac{3+j}{2} \right) \sum_{s=1}^{{N_{K_a}}} {\hat{n}_{a_s}}  \sum_{r=s}^{{N_{K_a}}} {\hat{n}_{a_r}} \frac{1}{{{\hat{v}}_{{ath}_r}}^7} \frac{1}{{\hat{v}_{{ath}_s}}^7} \left [2^{-(3+j)/2} ({{\hat{v}}_{{ath}_r}})^{4+j} - 
      \right. \\& \left.
    \left({{\hat{v}}_{{ath}_r}} \right)^{3+j} \frac{{\hat{v}_{{ath}_s}}^2-{{\hat{v}}_{{ath}_r}}^2}{{\hat{v}_{{ath}_s}}} {\mathrm{_2 F_1}} \left(\frac{1}{2},\frac{3+j}{2},\frac{3}{2},-\frac{{{\hat{v}}_{{ath}_r}}^2}{{\hat{v}_{{ath}_s}}^2}\right) - 
    \right. \\& \left.
    {\hat{v}_{{ath}_s}} \left(\frac{1}{{{\hat{v}}_{{ath}_r}}^2} + \frac{1}{{\hat{v}_{{ath}_s}}^2}\right)^{-(3+j)/2}
      \right],  \quad j \ge -2
      ~. \label{Rhaaj000D1V}
  \end{aligned}
\end{eqnarray}

The combination of 
the transport equation\EQ{dtMho}, kinetic dissipative force closure relation\EQ{Rhj0a0D1V},
characteristic parameter equations\EQ{CPEs9Ms0D1V} and the analytical expression of the normalized kinetic dissipative
force represented by Eqs.\EQ{Rhabj000D1V}-\EQo{Rhaaj000D1V} constitutes a set of nonlinear equations. These nonlinear equations,
for the situation when velocity space exhibits spherical symmetry without shell structure, will be referred as kinetic moment-closed model. The moment-closed model for homogeneous plasmas is a relaxation model.
The flowchart to solve this nonlinear model is provided in Appendix\APP{Procedure}.

Specially, the transport equations of mass density\EQ{dtna}, momentum\EQ{dtIa} and energy\EQ{dtKa} of spices $a$ will be:
\begin{eqnarray}
    {\frac{\partial}{\partial t}} \rho_a \left(t \right)  &=& {\frac{\partial}{\partial t}} I_a \left(t \right) \ = \ 0  \label{dtna0D1V}
\end{eqnarray}
and 
\begin{eqnarray}
    {\frac{\partial}{\partial t}} K_a \left(t \right) &=& \frac{1}{2}  \rho_a {{v}_{ath}} ^{2} \left({\frac{n_b}{{{v}_{bth}}^3}} {\Gamma_{ab}} {\hat{\mathcal{R}}}{_{ab}}_{2,0}  + {\frac{n_a}{{{v}_{ath}}^3}} {\Gamma_{aa}} {\hat{\mathcal{R}}}{_{aa}}_{2,0} \right) . \label{dtKa0D1V}
\end{eqnarray}
Applying the relation, $K_a = \frac{3}{2} n_a T_a$, and mass conservation represented by Eq.\EQ{dtna0D1V} results in a temperature relaxation equation:
\begin{eqnarray}
    {\frac{\partial}{\partial t}} T_a \left(t \right) &=&- {\nu_T^{a}} T_a, \label{dtTa0D1V}
\end{eqnarray}
where the characteristic frequency of temperature relaxation,
\begin{eqnarray}
    {\nu_T^{a}} \left(t \right) &=&- \frac{2}{3} \left({\frac{n_b}{{{v}_{bth}}^3}} {\Gamma_{ab}} {\hat{\mathcal{R}}}{_{ab}}_{2,0}  + {\frac{n_a}{{{v}_{ath}}^3}} {\Gamma_{aa}} {\hat{\mathcal{R}}}{_{aa}}_{2,0} \right) ~.\label{nuTtab0D1V}
\end{eqnarray}

Due to the numerous advantages of KFE, the current relaxation model, serving as a moment approach of NLVFP, also offers several benefits:
\begin{itemize}
    \item[I)] The relaxation model explicitly provides the analytical forms of nonlinear kinetic dissipative closure relations\EQ{Rhj0a0D1V} based on arbitrary order kinetic dissipative forces\EQ{Rhabj000D1V}-\EQo{Rhaaj000D1V}. 
    \item[II)] The relaxation model is founded on the conserved moments and high-order kinetic moments to describe the system evolution\EQ{dtMho}. Therefore, it is more suitable for constructing numerical algorithms with high-order moment convergence. 
    \item[III)] The relaxation model adaptively determines the optimal number of sub-distribution functions at each time level, based on the CPEs\EQ{CPEs9Ms0D2V}, enabling a more precise solution of the corresponding VFP equation. 
\end{itemize}

The kinetic effects are depicted by the higher-order kinetic moments in this relaxation model. The
advantages of this model make it suitable for steady-state homogeneous fusion plasmas which may be
far from thermodynamic equilibrium state. Obviously, it is a well-established fact that the presence of
fast $\alpha$ particles in fusion plasmas leads to abundant nonlinear interactions between itself with electrons
and fusion fuel ions. This is particularly evident when considering multi-time scales relaxation problems
of homogeneous plasmas, which are typically nonlinear, dominating by lower-order kinetic moments
and influencing by higher-order kinetic moments. 
The proposed model can easily capture the inherent
nonlinearity of these plasmas system with higher efficiency and stability.

This relaxation model, as a moment approach in NLVFP\cite{wang2024higherorder}, is a further advancement of the meshfree approach\cite{wang2024Aconservative}, which is based on expanding the distribution function in spherical harmonics in angle
coordinate and in King basis in speed coordinate of velocity space. Both of these higher-order moment
convergent method in NLVFP can be extended to two-dimension or three-dimension velocity space by
introducing new special functions, namely (associate) King function\cite{wang2024Aconservative} and R function\cite{wang2024General}. The extended
transport equations can be applied for conventional magnetic confinement fusion (MCF) simulations
by including the terms of spatial convection and mean-field effects in VFP equation. As noted by Bell\cite{Bell2006}, the rotating effect of a magnetic field is easily and accurately represented in SHE approach. As
a consequence, coupling SHE with KFE, this moment approach has significant advantages for studying
the fusion plasmas with strong magnetic fields.

\subsubsection{Special case: Two-temperature thermal equilibrium model}
\label{Two-temperature thermal equilibrium model}

The numbers of sub-distribution are both equal to 1, ${N_{K_a}}={N_{K_b}} \equiv 1$, when the two species are in thermal equilibrium at different temperatures. Consequently, ${\hat{n}_{a_r}} = {\hat{n}_{b_s}} \equiv 1$ and ${{\hat{v}}_{{ath}_r}} = {{\hat{v}}_{{bth}_s}} \equiv 1$, leading to the simplification of Eq.\EQ{Rhabj000D1V}, reads: 
      \begin{eqnarray}
            {\hat{\mathcal{R}}}{_{ab}}_{j,0} \left(t \right) &=&
          \left \{
            \begin{aligned}
            & 0, \quad j = 0, \label{Rh000NK1}
            \\
            & \frac{C_j^{R}}{2} \left[\left(\frac{1}{{{v}_{abth}}} + {{v}_{abth}} \right) \arctan ({{v}_{abth}}) - 1 \right], \quad j = 1, \label{Rh100NK1}
            \\
            & C_j^{R} \left[1 + \sum_{k=2}^{j/2} ({{v}_{abth}})^{2k} {\boldsymbol{c}}_j[k] \right] , \quad j \in 2\mathbb{N}^+, \label{Rhj2N2}
            \\
            & C_j^{R} \left[{\mathrm{_2 F_1}} \left(- \frac{j}{2},1,\frac{3}{2}, - {{v}_{abth}}^2 \right)-1 \right], \quad j \in 2\mathbb{N}^+ + 1 ~. \label{Rhj2N1}
            \\
            \end{aligned} \label{Rhabj00nuTs119Ms0D2V}
          \right.
      \end{eqnarray}
      
Operator ${\boldsymbol{c}}_j\left[k\right]$ in Eq.\EQ{Rhj2N2} represents the $k^{th}$ element of the vector ${\boldsymbol{c}}_j$, and satisfies the following recursive relationship:
\begin{eqnarray}
    {\boldsymbol{c}}_j[k]&=&\frac{j - 2 k}{2 k + 3} {\boldsymbol{c}}_j[k-1], \ {\boldsymbol{c}}_j[1] = 1, \ 2 \le k \le j/2 ~. \label{Cj2NMatrix}
\end{eqnarray}
Parameter
      \begin{eqnarray}
            C_j^{R} &=&
          \left \{
            \begin{aligned}
            &  \frac{4 \pi}{\pi^{3/2}} \frac{j(j+1)!!}{3 \sqrt{2^j}} \frac{1}{{{v}_{abth}}^2} \frac{1 - m_M {{v}_{abth}}^2}{\sqrt{\left(1+{{v}_{abth}}^2 \right)^{j+1}}} , j \in 2\mathbb{N}^+, \label{Cj2N9Ms0D2V}
            \\
            &  \frac{8}{\pi} \frac{\left(j/2+1/2 \right)!}{{{v}_{abth}}^4} \frac{1 - m_M {{v}_{abth}}^2 }{\sqrt{\left(1+{{v}_{abth}}^2 \right)^{j+1}}}, j \in 2\mathbb{N}^+ + 1 ~. \label{Cj2N19Ms0D2V}
            \\
            \end{aligned} \label{CjR9Ms0D2V}
          \right.
      \end{eqnarray}
 Eq.\EQ{Rhabj00nuTs119Ms0D2V} reveals that the arbitrary order normalized kinetic dissipative force solely depends on $m_M$ and ${{v}_{abth}}$, indicating that the high-order normalized kinetic dissipative force is not an independent quantity when the two species are in thermal equilibrium respectively.

Similarly, Eq.\EQ{Rhaaj000D1V} reduces to be:
\begin{eqnarray}
    {{{\hat{\mathcal{R}}}{_{aa}}}{_{j,0}}} \left(t \right) & \equiv &  0,  \quad j \ge -2
      ~. \label{Rhaaj000D1V0}
\end{eqnarray}
In other words, any order normalized kinetic moments during self-collision process remain constant over time when the distribution function of species $a$ is in thermodynamic equilibrium. In this case, the transport equation\EQ{dtMho} will be:
  \begin{eqnarray}
      {\frac{\partial}{\partial t}} {\mathcal{M}_{j,l}} \left(t \right) &=& \delta_l^{0} \rho_a \left({{v}_{ath}} \right)^j {\frac{n_b}{{{v}_{bth}}^3}} {\Gamma_{ab}} {{\hat{\mathcal{R}}}{_{ab}}_{j,l}}  ,  \label{dtMj000D1V}
  \end{eqnarray}
where function ${{\hat{\mathcal{R}}}{_{ab}}_{j,l}}$ satisfies Eq.\EQ{Rhabj00nuTs119Ms0D2V}. 
 
\subsubsection{Special case: Braginskii heat transfer model}
\label{Braginskii model}

In particular, the $(2,0)^{th}$-order transport equation, when the two species are in thermodynamic equilibrium at different temperatures, will be:
  \begin{eqnarray}
      {\frac{\partial}{\partial t}} \mathcal{M}_{2,0} \left(t \right) &=& \frac{4 \pi}{\pi^{3/2}} \rho_a {{v}_{ath}}^2 {\frac{n_b}{{{v}_{bth}}^3}} {\Gamma_{ab}} \frac{1 - m_M {{v}_{abth}}^2}{{{v}_{abth}}^2 \sqrt{\left(1+{{v}_{abth}}^2 \right)^3}}  ~.  \label{dtMh2000D1V}
  \end{eqnarray}
Substituting Eq.\EQ{KaMh} into the above equation yields:
  \begin{eqnarray}
      {\frac{\partial}{\partial t}} T_a \left(t \right) &=& \frac{1}{3} \frac{4 \pi}{\pi^{3/2}} m_a {{v}_{ath}}^2 {\frac{n_b}{{{v}_{bth}}^3}} {\Gamma_{ab}} \frac{1 - m_M {{v}_{abth}}^2}{{{v}_{abth}}^2 \sqrt{\left(1+{{v}_{abth}}^2 \right)^3}}  ~.  \label{dtTa0D1V0}
  \end{eqnarray}
Substituting the expression of ${\Gamma_{ab}}$ into above equation and applying the following relation,
  \begin{eqnarray}
      \frac{1 - m_M {{v}_{abth}}^2}{{{v}_{abth}}^2 \sqrt{\left(1+{{v}_{abth}}^2 \right)^3}} 
      & = & \frac{2}{2^{3/2}} \frac{{{v}_{bth}}^3 \left(m_a m_b \right)^{3/2}}{m_b {{v}_{ath}}^2} \frac{T_b - T_a}{\left(m_a T_b+m_b T_a \right)^{3/2}} , 
  \end{eqnarray}
we can obtain:
  \begin{eqnarray}
      {\frac{\partial}{\partial t}} T_a \left(t \right) & = & \frac{16 \pi}{3 \sqrt{2 \pi}} \left(\frac{q_e^2}{4\pi \varepsilon_0} \right)^2 \frac{\sqrt{m_a m_b} \left(Z_a Z_b \right)^2 n_b {\ln{ \Lambda_{ab}}}}{\left(m_a T_b+m_b T_a \right)^{3/2}} \left (T_b - T_a \right)  ~.
  \end{eqnarray}
 
The above equation can be simplified and expressed as follows:
  \begin{eqnarray}
      {\frac{\partial}{\partial t}} T_a \left(t \right) &=& - {\nu_T^{ab}} \left (T_a - T_b \right) ~.  \label{dtTa9Ms0D2V}
  \end{eqnarray}
The characteristic frequency of temperature relaxation is consistent with the result obtained by Huba\cite{Huba2011}, which can be expressed as:
\begin{eqnarray}
    {\nu_T^{ab}} &=& \frac{8\sqrt{2\pi}}{3} \left(\frac{q_e^2}{4\pi \varepsilon_0} \right)^2 \frac{\sqrt{m_a m_b} \left(Z_a Z_b \right)^2 n_b}{\left(m_a T_b + m_b T_a \right)^{3/2}} {\ln{ \Lambda_{ab}}}  ~. \label{nuTab00D1V}
\end{eqnarray}
In above equation, $q_e$ is the charge of the positron; $Z_a$ and $Z_b$ are the particle charge number of species $a$ and $b$, respectively. 
Eq.\EQ{dtTa9Ms0D2V} exhibits the same form as the Braginskii heat transfer model\cite{Braginskii1958}, which serves a benchmark model for VFP simulation\cite{Taitano2016}. We denote this heat transfer model as zeroth-order Braginskii model. This model ignores the deviation of background specie distribution function from the Maxwellian one under the near-equilibrium assumption. However, this is a nature consequence of our relaxation model represented by Eq.\EQ{dtTa0D1V} when both the distribution functions during Coulomb collision are Maxwellian.

\subsubsection{Special case: Thermodynamic equilibrium model}
\label{Thermodynamic equilibrium model}

Furthermore, in the state of thermodynamic equilibrium for the plasmas system, where the numbers of sub-distributions are both equal to 1 and both species have the same temperature ($T_a=T_b$, especially when $m_M = {{v}_{abth}} =1$), it follows that the factor in Eq.\EQ{CjR9Ms0D2V} becomes $1 - m_M {{v}_{abth}}^2 \equiv 0$. 
Consequently, Eq.\EQ{Rhabj00nuTs119Ms0D2V} can be expressed as follows: 
\begin{eqnarray}
    {{{\hat{\mathcal{R}}}{_{ab}}}{_{j,0}}} \left(t \right) &=& 0 , \quad \forall j ~. \label{Rhj0ab0D1VTaTb}
\end{eqnarray}

Substituting Eq.\EQ{Rhj0ab0D1VTaTb} into Eq.\EQ{dtMj000D1V} yields the transport equation for homogeneous plasmas system in thermal equilibrium, reads:
  \begin{eqnarray}
      {\frac{\partial}{\partial t}} {{\mathcal{M}}{_{j,0}}} \left(t \right) & \equiv & 0 , \quad \forall j ~.  \label{dtMj000D1V0}
  \end{eqnarray}
In other words, if both species are in thermodynamic equilibrium and have the same temperature during Coulomb collision process, any order of the system's kinetic moment does not spontaneously change with time.

\section{Conclusion}
\label{Conclusion}

It has been demonstrated that a relaxation model is obtained when the velocity space exhibits spherical symmetry without shell structures.
This model comprises a set of transport equations of arbitrary order (include density, momentum, and energy) based on Maxwellian mixture model. These results are typically presented in closed form in term of Gauss hypergeometric2F1 functions. 
Furthermore, it has been demonstrated that 
our relaxation model encompasses specific instances such as the two-temperature thermal equilibrium model, zeroth-order Braginskii heat transfer model, and thermodynamic equilibrium model.

It is important to note that our article focuses on proposing a mixture model based on the finitely distinguishable independent feature hypothesis rather than traditionally employed near-equilibrium 
assumption. 
We have derived the relaxation model for a two-species plasmas with spherically symmetric velocity space using FPRS collision operator. The results accurately capture both near-equilibrium and far-from-equilibrium states for spherically symmetric plasmas system. 
The analysis of scenarios within more general velocity space, including those related to axisymmetric
systems, has been accomplished and will be published in the future.
These findings will serve as valuable benchmarks for nonlinear statistical physics applications such as fusion plasmas and solar plasmas.

\section{Acknowledgments}
\label{Acknowledgments}

We would like to thank Yi-Feng Zheng, Zhi-Hui Zou, Jian Zheng, Zhe Gao and Meng-ping Zhang for useful discussions. This work is supported by the Strategic Priority Research Program of the Chinese Academy of Sciences (Grant No. XDB0500302), and the Laoshan Laboratory (No.LSKJ202300305).


\appendix    

\begin{appendices}

\section{Flowchart to solve the relaxation model based on MMM}
\label{Procedure}

The relaxation model based on MMM
for scenario with shell-less spherical symmetric velocity
space, described in detail in Sec.\SEC{Kinetic moment-closed model based on MMM}, generally consists of a set of nonlinear equations that can only be solved numerically. In this paper, we just present the flowchart to solve these nonlinear equations, which is provided in Fig.\FIG{flowchart}. The transport equation represented by Eq.\EQ{dtMho} can be solved by a Runge-Kutta solver, i.e., trapezoidal\cite{Rackauckas2017} scheme which is a second-order implicit method. 
A general relaxation model based on KMM0 for scenario with general spherical symmetric velocity space has been provided in Ref.\cite{wang2024General} and will be published in the near future.

\tikzstyle{startstop} = [rectangle, rounded corners, minimum width = 2cm, minimum height=1cm,text centered, draw = black, fill = red!40]
\tikzstyle{io} = [trapezium, trapezium left angle=70, trapezium right angle=110, minimum width=2cm, minimum height=1cm, text centered, draw=black, fill = blue!40]
\tikzstyle{process} = [rectangle, minimum width=3cm, minimum height=1cm, text centered, draw=black, fill = yellow!50]
\tikzstyle{decision} = [diamond, aspect = 3, text centered, draw=black, fill = green!30]
\tikzstyle{arrow} = [->,>=stealth]
\begin{figure}[htp]
    \centering
    \thispagestyle{empty}
    \begin{tikzpicture}[node distance=2cm]
\node (start) [startstop, xshift= -0.0cm] {Input ${f_0}(v,t_0)$\EQ{flm} for all species};
\node (pro1) [process, below of=start, yshift= +0.56cm] {Initial $3 {N_{K_a}}({t_k})$ kinetic moments $M_{j,0}(t=0) = {{\mathcal{M}}{_{j,0}}}(t=0)$\EQ{Mjl3D2V} and $\rho_a({t_k})$\EQ{naM}, ${{v}_{ath}}({t_k})$\EQ{vathM}};
\node (pro12) [process, below of=pro1, yshift= +0.56cm] {Solving the CPEs\EQ{CPEs9Ms0D1V}};
\node (pro2) [process, below of=pro12, yshift= +0.56cm] {Updating characteristic parameters ${\hat{n}_{a_r}}({t_k})$, ${\hat{u}_{a_r}}({t_k})$ and ${{v}h_{{ath}_r}}({t_k})$};
\node (pro3) [process, below of=pro2, yshift= +0.56cm] {Update ${{{\hat{\mathcal{R}}}{_{ab}}}{_{j,0}}}({t_k})$ according to Eq.\EQ{Rhabj000D1V} and ${{{\hat{\mathcal{R}}}{_{aa}}}{_{j,0}}}({t_k})$ according to Eq.\EQ{Rhaaj000D1V}};
\node (pro4) [process, below of=pro3, xshift= -0cm, yshift= +0.56cm] {Compute the kinetic dissipative force closure relation\EQ{Rhj0a0D1V} to obtain ${{\hat{\mathcal{R}}}{_{j,0}}} ({t_k})$};
\node (pro5) [process, below of=pro4,  yshift= +0.56cm] {Update $M_{j,0}({t_{k+1}})$\EQ{Mjl3D2V} according to Eq.\EQ{Mjltk1} and renew $\rho_a({t_{k+1}})$\EQ{naM}, ${{v}_{ath}}({t_{k+1}})$\EQ{vathM}};
\node (pro6) [process, below of=pro5,  yshift= +.56cm] {Renew ${N_{K_a}}({t_{k+1}})$, update ${{\mathcal{M}}{_{j,0}}}({t_{k+1}})$ and solve the new well-posed CPEs\EQ{CPEs9Ms0D1V}};
\node (dec1) [decision, below of=pro6, yshift= +0.16cm] {$\sum_j \delta {{\mathcal{M}}{_{j,0}}}({t_{k+1}}) \le Rtol$};
\node (pro7) [process, below of=dec1,  yshift= +.16cm] {Update $d{N_{K_a}}({t_{k+1}})$};
\node (pro8) [startstop, below of=pro7, yshift= +0.56cm] {Output $\rho_a({t_k})$, ${{v}_{ath}}({t_k})$ and ${\hat{f}_0}(v,{t_k})$\EQ{fhofhor}};
\draw [arrow](start) -- (pro1);
\draw [arrow](pro1) -- (pro12);
\draw [arrow](pro12) -- (pro2);
\draw [arrow](pro2) -- (pro3);
\draw [arrow](pro3) -- (pro4);
\draw [arrow](pro4)  -- (pro5);
\draw [arrow](pro5)  -- (pro6);
\draw [arrow](pro6) -- (dec1);
\draw [arrow](dec1) -- node[above]{Y} ++(- 7.5, 0) -- ++(0, 5.5) |- (pro2);
\draw [arrow](dec1) --node[right]{N} (pro7);
\draw [arrow](pro7) -- ++(+ 7.5, 0) -- ++(0, 2.5) |- (pro6);
\draw [arrow](dec1) -- ++(- 7.5, 0) -- ++(0, -3) |- (pro8);
    \end{tikzpicture}
    
    \caption{Flowchart of the kinetic moment-closed model for plasma when velocity space exhibits spherical symmetry without shell structure}
    \label{flowchart}
\end{figure}
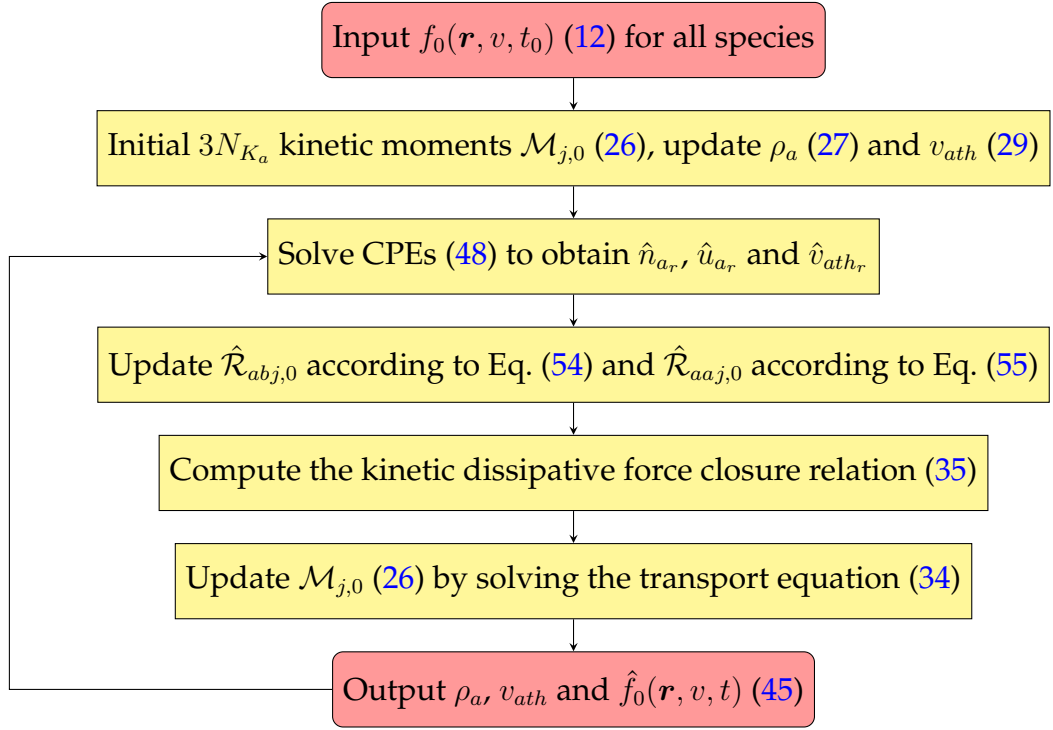

\section{Convergence of KMM0 and MMM}
\label{Convergence of KMM0 and MMM}

Convergence of KMM0 and MMM can be proved based on the Wiener's Tauberian theorem.
Wiener's Tauberian theorem is a set of important theorems about smooth function approximation that was proposed by Norbert Wiener in 1932 \cite{Wiener1932}, which will be quoted as follows:

\

\begin{thm}{}(Wiener's Tauberian theorem) \label{Wiener's Tauberian theorem}

    Let $f \in L^1(\mathbb{R})$ be an integrable function. The span of translations  $g_a(x) = g(x + a)$ is dense in  $L^1(\mathbb{R})$ if and only if the Fourier transform of function $f$ has no real zeros.

\end{thm}

\

The Fourier transform of Gaussian function is still a Gaussian function. Therefore, the Gaussian function is obviously a dense function in the Euclidean space. Gaussian function serves as a commonly employed non-orthogonal basis for two primary reasons: Firstly, many natural statistical systems are independent and identically distributed systems that often adhere to the central limit theorem \cite{Rosenblatt1956}; Secondly, Gaussian functions offer computational simplicity. By selecting the basis $g(v)$ as a Gaussian function to approximate the distribution function $f(v)$, i.e., $F(v)=\sum_{s=1}^{N} w_s g(v+c_s)$, where $w_s$ is the weight of basis function. This form represents the 
one-dimensional Gaussian mixture model \cite{Banerjee2013} (GMM)
with identical  expectation.

GMM with different expectations and deviations is represented by approximating the function $f(v)$ with a series of scaled and translated Gaussian functions, $g\left[\left(v+c_{s,k}\right)/\sigma_{s,k}\right]$. With $N$ expectations and $N$ deviations, this approximation can be expressed as:
\begin{eqnarray}
    F(v)  &=&  \sum_{k=1}^{N} \sum_{k=1}^{N} w_{s,k}g\left[\left(v+c_{s,k}\right)/\sigma_{s,k}\right] ~. \label{FvNkNk}
\end{eqnarray}
Above equation consists of a total of $N \times N$ Gaussian functions, which can be reduced by utilizing optimization algorithms, such as the expectation-maximization\cite{Wynne2021} (EM) method. After obtaining an optimized set of parameters $\{w_{s,k}, c_{s,k}, \sigma_{s,k}\}$ where $s=k$,
Eq.\EQ{FvNkNk} can be expressed as:
\begin{eqnarray}
    F(v)  &=& \sum_{k=1}^{N} w_{k}g\left[\left(v+c_{k}\right)/\sigma_{k}\right] ~. \label{FvNk}
\end{eqnarray}
Specifically, when all deviations are zero ($c_k \equiv 0,\forall k$), Eq.\EQ{FvNk} will be:
\begin{eqnarray}
    F(v) &=& \sum_{k=1}^{N} w_k g\left(v/\sigma_k\right) ~. \label{FvNku0}
\end{eqnarray}
Eq.\EQ{FvNku0} is the form of MMM\EQ{fhofhor} given in Sec.\SEC{Finitely distinguishable independent features hypothesis}. 
Function ${\mathcal{K}_0}$\EQ{K0} is the form of Gaussian function in spherical coordinate system when the velocity space is spherically symmetric, which can be obtained by employing the spherical harmonic expansion\cite{Arfken1971}. Hence, KMM0\EQ{fhoKo} will be convergent when the velocity space exhibits spherical symmetry.


\end{appendices}
 
\end{spacing}
 
\begin{spacing}{0.5}  
 
\
\


\end{spacing}


\bibliographystyle{iopart-num.bst}
\bibliography{Plasma}

\end{document}